\documentclass[aps,prl,twocolumn,superscriptaddress]{revtex4-1}
\usepackage{amsmath,amssymb,graphicx,epstopdf,bm,hyperref}

\begin{document} 
\bibliographystyle{apsrev4-1}

\title{\textit{Extreme thermodynamics} with polymer gel tori: harnessing thermodynamic instabilities to induce large-scale deformations}

\author{Ya-Wen Chang}\thanks{These two authors contributed equally}
\affiliation{School of Physics, Georgia Institute of Technology, Atlanta, Georgia 30332, USA}
\affiliation{Present address: Department of Chemical Engineering, Texas Tech University, Lubbock, TX 79409, USA}
\author{Michael S. Dimitriyev}\thanks{These two authors contributed equally}
\affiliation{School of Physics, Georgia Institute of Technology, Atlanta, Georgia 30332, USA}
\author{Anton Souslov}
\affiliation{School of Physics, Georgia Institute of Technology,
Atlanta, Georgia 30332, USA}
\affiliation{Leiden Institute of Physics, Leiden University, Niels Bohrweg 2, 2333 CA Leiden, Netherlands}
\affiliation{The James Franck Institute, University of Chicago, 929 E 57th Street, Chicago, IL 60637, USA}
\author{Svetoslav V. Nikolov}
\affiliation{George W. Woodruff School of Mechanical Engineering, Georgia Institute of Technology, Atlanta, Georgia 30332, USA}
\author{Samantha M. Marquez}
\affiliation{Branford College, Yale University, New Haven, Connecticut 06520, USA}
\author{Alexander Alexeev}
\affiliation{George W. Woodruff School of Mechanical Engineering, Georgia Institute of Technology, Atlanta, Georgia 30332, USA}
\author{Paul M. Goldbart} 
\affiliation{School of Physics, Georgia Institute of Technology, Atlanta, Georgia 30332, USA}
\author{Alberto Fern\'{a}ndez-Nieves}
\affiliation{School of Physics, Georgia Institute of Technology, Atlanta, Georgia 30332, USA}

\date{\today}

\begin{abstract}
When a swollen, thermoresponsive polymer gel is heated in a solvent bath, it expels solvent and deswells.
When this heating is slow, deswelling proceeds homogeneously, as observed in a toroid-shaped gel that changes volume whilst maintaining its toroidal shape.
By contrast, if the gel is heated quickly, an impermeable layer of collapsed polymer forms and traps solvent within the gel, arresting the volume change.
The ensuing evolution of the gel then happens at fixed volume, leading to phase-separation and the development of inhomogeneous stress that deforms the toroidal shape.
We observe that this stress can cause the torus to buckle out of the plane, via a mechanism analogous to the bending of bimetallic strips upon heating.
Our results demonstrate that thermodynamic instabilities, i.e.~phase transitions, can be used to actuate mechanical deformation in an \textit{extreme thermodynamics} of materials.
\end{abstract}


\maketitle

\begin{figure}
\includegraphics[width=8.6cm]{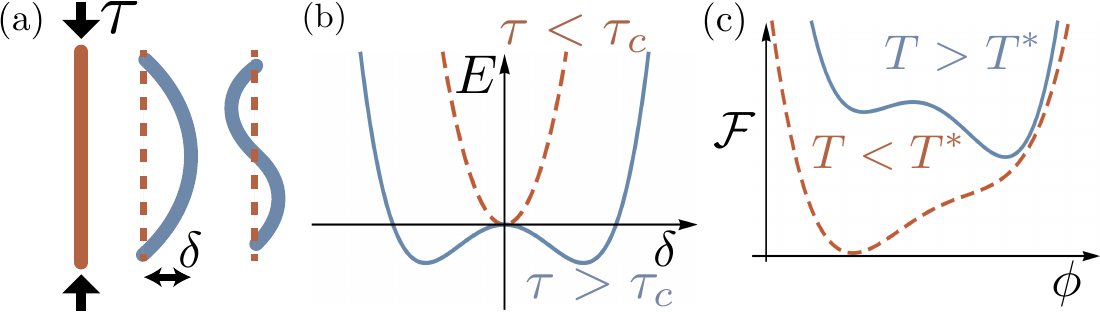}
\caption{\label{fig1} (a) Elastic rod that is compressed at its ends by a tension $\tau$. Left: straight rod; right: two examples of buckled rods. (b) Total energy of a compressed rod as a function of deflection for values of tension $\tau$ less than (dashed) and greater than (solid) a critical tension $\tau_c$. (c) Free-energy density of a polymer gel as a function of polymer volume fraction $\phi$ for temperatures below a transition temperature $T^*$, where the gel is in the swollen phase at low $\phi$, and above $T^*$, where it can be forced into a phase coexistent state.} 
\end{figure}

The term ``extreme mechanics" is often used in reference to mechanical structures with prescribed instabilities that enable large deformations and configurations that are hard to achieve by other means \cite{Reis2015}.
An example of this is Euler buckling, which refers to the case of a straight, slender, homogeneous elastic rod that is compressed at its ends by an applied stress [Fig.~\ref{fig1}(a)] \cite{LandauLifshitz1986}.
Below a critical stress, $\tau_c$, there is a stable energy minimum corresponding to the deflectionless equilibrium configuration of a straight rod [Fig.~\ref{fig1}(b), dashed curve].
In contrast, above $\tau_c$, the energy minimum becomes a maximum and the straight rod configuration becomes unstable, with two new minima describing the stable, bent configuration of the rod [Fig.~\ref{fig1}(b), solid curve]; this deformed state is thus achieved via a mechanical instability above $\tau_c$.

Experimentally, shape actuation is often realized with polymeric materials, such as polymer gels, which are crosslinked polymer networks immersed in a solvent \cite{ShibayamaTanaka1993}.
These respond to external stimuli by swelling or deswelling and equilibrate when the total free-energy, consisting of a polymer-solvent mixing contribution and the entropic elasticity of the polymer network, is minimized \cite{flory1953principles}.
In a so-called \textit{thermoresponsive gel}, the interplay between these two contributions to the free energy can be adjusted via temperature.
Interestingly, if there are inhomogeneities in the polymer distribution within the gel, striking swelling patterns \cite{Arifuzzaman} can be achieved; these are oftentimes similar to the topographical features observed in soft tissues~\cite{Trujillo2008,Breid,Hohlfeld_sulcus}.
This strategy has also proven useful in the design of tunable surface patterns~\cite{Guvendiren} and self-folding origami~\cite{Stoychev}. 
Importantly, in all these instances, the gel swells \emph{quasistatically} and is thus equilibrated with the surrounding solvent bath throughout the process.

\begin{figure*}
\includegraphics{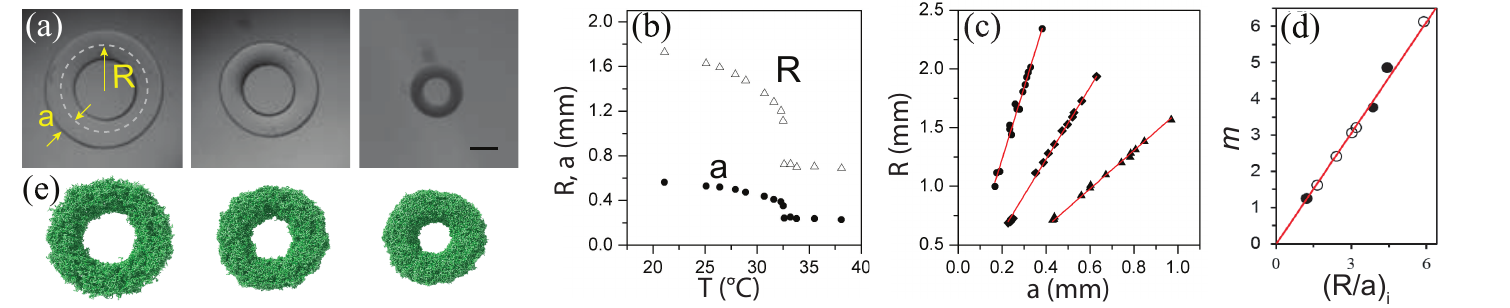}
\caption{\label{fig2} (a) Quasistatic deswelling of a toroidal gel equilibrated at 25.1, 30.7, 33.5 $^{\circ}$C. Scale bar: 1mm. (b) Temperature-dependence of the ring radius, $R$, and the tube radius, $a$, for a torus of initial aspect ratio $(R/a)_i$ = 3.0. (c) $R$ vs~$a$ for tori with an $(R/a)_i$ of ($\blacktriangle$) 1.6, ($\blacklozenge$) 3.0, and ($\bullet$) 5.9 undergoing quasistatic deswelling. The solid lines are linear fits to the data. The intercepts of the fits are, from top to bottom (in mm): $(0.02 \pm 0.06)$ mm, $(-0.037 \pm 0.015)$ mm, and $(0.042 \pm 0.011)$ mm; these are all close to zero, consistent with deswelling happening at constant $\xi$. The slopes $m$ of these fits are shown in (d) as a function of $(R/a)_i$. The closed symbols are the results obtained in computer simulations. The solid line corresponds to $m = (R/a)_i$. (e) Simulation snapshots of a toroidal gel that is deswelling quasistatically.} 
\end{figure*}

However, polymer gels can also exhibit discontinuous phase transitions between polymer-solvent mixed and segregated phases, corresponding to swollen and deswollen states. Furthermore, they can also exhibit phase coexistence where different parts of the gel are either solvent-rich or solvent-poor  \cite{ShibayamaTanaka1993,Hirotsu1993}.
In thermoresponsive gels below a threshold temperature, $T^*$, the system is in an equilibrium swollen state, where the free energy is minimum [Fig.~\ref{fig1}(c), dashed curve].
In contrast, above $T^*$, the gel can exhibit phase coexistence and be characterized by a free energy with two minima [Fig.~\ref{fig1}(c), solid curve].
Importantly, due to the gel's shear rigidity, this last equilibrium arrangement of coexistent phases must additionally minimize the free-energy cost associated with the inhomogeneous distribution of the polymer network.
Swelling equilibria thus depend on and influence the shape of the gel; the order parameter associated to the phase transition then couples to the shape, potentially affecting it in what we could call an \textit{extreme thermodynamics} of materials.
Unlike the mechanical case of Euler buckling, in this case, a thermodynamic instability is exploited to achieve large-scale material deformations.

In this Letter, we explore this idea using thermoresponsive gels made of poly-N-isopropylacrilamide (pNIPAM) and shaped as a toroid.
First, we demonstrate the actuation of volume changes at fixed toroid shape.
Next, we discuss our observations that after rapid heating, the toroid undergoes large shape changes and buckles out of the plane.
We find that the toroid undergoes internal phase-separation at constant volume, leading to a \emph{polarized} arrangement of solvent and polymer within its cross-section that results in a substantial internal stress difference.
Through simulation and analytical modeling, we demonstrate that the observed arrangement is responsible for the toroid's buckling, confirming the notion of \textit{extreme thermodynamics} as a means to achieve shape actuation.

We fabricate toroidal gels by first forming toroidal droplets of a precursor NIPAM solution, which is then UV-polymerized \cite{sup-mat,chang2015biofilm,Pairam3}. When heated past the lower critical solution temperature (LCST), pNIPAM gels enter a deswollen, polymer-rich phase, characterized by a small volume. 
Snapshots of the quasistatic evolution of a toroidal gel are shown in Fig.~\ref{fig2}(a). 
Both the ring radius, $R$, and the tube radius, $a$, decrease with increasing temperature, as shown in Fig.~\ref{fig2}(b). 
The rate of decrease is highest at ~32.5$^{\circ}$C, which corresponds to the LCST of pNIPAM~\cite{schild1992poly}.  
Above this temperature, both $R$ and $a$ remain essentially constant, as also shown in Fig.~\ref{fig2}(b); at these temperatures the gel is deswollen and optically opaque, as seen in the rightmost image in Fig.~\ref{fig2}(a).

\begin{figure*}
\includegraphics{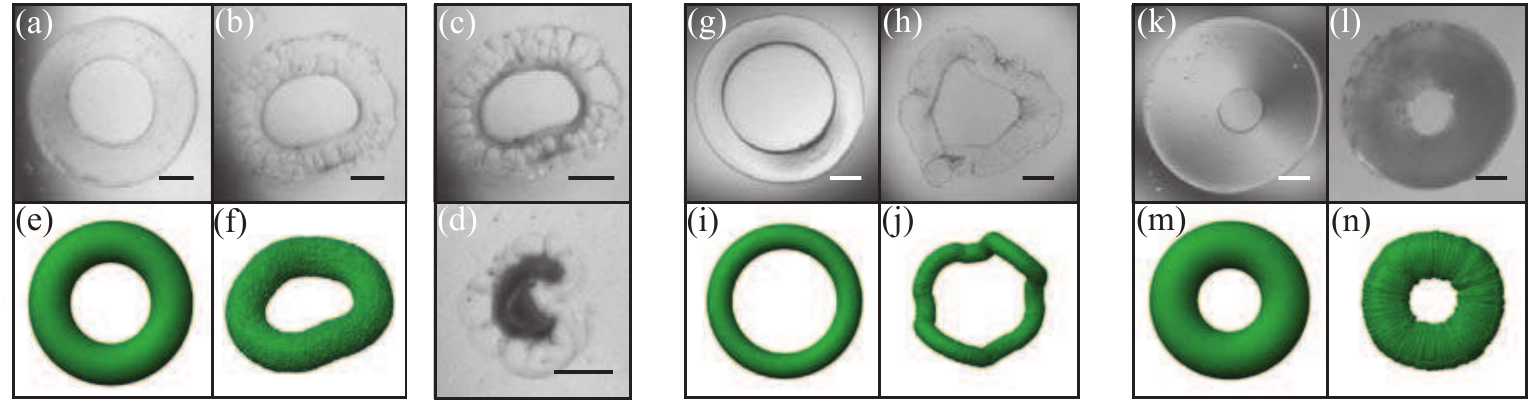}
\caption{\label{fig3} Evolution of toroidal gels after rapid heating. 
The experimental image pairs (a,b), (g,h), and (k,l) are taken (a,g,k) 0s, (b) 104s, (h) 158s, and (l) 201s after heating. Pairs (e,f), (i,j), and (m,n) are simulations for gel shells; images on the left correspond to the initial state, while images on the right correspond to the final state. The initial aspect ratio of the tori are: (a) 3.3, (e) 3.0, (g) 4.8, (i) 5.3, 
(k) 1.7, and (m) 2.4. Images (c,d) correspond to the long-time evolution of the toroidal gel in (a,b). (c) is 3 min and (d) is 10 mins after the abrupt temperature change. Scale bars: 1mm.}
\end{figure*}

Since the gel remains isotropic and homogeneous during the quasistatic heating process, any change in the polymer matrix brought about by changes in $\phi$ must occur uniformly throughout the gel.  
Thus, all macroscopic lengths are expected to rescale by the same amount, implying that the aspect ratio of the torus, $\xi \equiv R/a$, remains unchanged.
To test this, we plot $R$ as a function of $a$ for all tori as they deswell, and find that they are linearly related, as shown for three representative examples in Fig.~\ref{fig2}(c). 
We also find there is a one-to-one correspondence between the slopes, $m$, obtained from the linear fits of the data, and the aspect ratio of the tori measured before deswelling. 
This is shown in Fig.~\ref{fig2}(d), and confirms our expectations. 
We also perform dissipative particle dynamics (DPD) computer simulations to further test our results \cite{sup-mat}. 
Representative snapshots of a simulated gel as it deswells are shown in Fig.~\ref{fig2}(e). 
Consistent with the experimental results, $R$ is linearly related to $a$, with a slope that corresponds to the aspect ratio before deswelling; the associated data points are shown in  Fig.~\ref{fig2}(d) with closed symbols.

In striking contrast with these observations, when we rapidly raise the temperature from the swollen phase at $\sim 10^{\circ}$C to the deswollen phase at 40.0$^{\circ}$C, the gel buckles, adopting a ``Pringle\textsuperscript{TM}''-like shape, as shown for a torus with $\xi = 3.3$ in Fig.~\ref{fig3}(a,b). 
This state persists over time scales from minutes to hours, depending on the overall dimensions of the torus, and eventually evolves while developing other characteristic features, as shown in Figs.~\ref{fig3}(c,d). 
In spherical and cylindrical pNIPAM gels subjected to abrupt temperature changes, there is a ``plateau period'' over which the gel retains its original volume, followed by the formation of surface patterns~\cite{Matsuo} that are reminiscent of those we observe for tori at long-times [Figs.~\ref{fig3}(c,d)]. 
The origin of this non-quasistatic evolution is the formation of a deswollen, collapsed-polymer layer, leading to extremely slow deswelling of the bulk of the gel, which, as a result, essentially maintains a constant volume~\cite{Yoshida, Okano}. 
The long-time patterns seen in our toroidal gels suggest that a similar situation occurs in our case and that the evolution we observe after rapid heating essentially happens at constant volume; this is supported by the observation that the time over which the toroid buckles is much shorter than the plateau period.
Hence, after rapid heating, the gel is out of equilibrium with the solvent bath and is thus not constrained to maintain a constant osmotic pressure $\Pi_{\rm bath}$, but rather a constant volume.
In this situation, a swollen gel is not allowed to change its total polymer volume fraction.
However, since the swollen gel has been brought to a temperature above the LCST of pNIPAM, the homogeneously mixed state of the gel becomes unstable to separation into solvent-rich and solvent-poor regions.
We then postulate that the shape transformation observed in experiment is due to this phase-separation at constant volume.

\begin{figure}
\includegraphics[width=8.6cm]{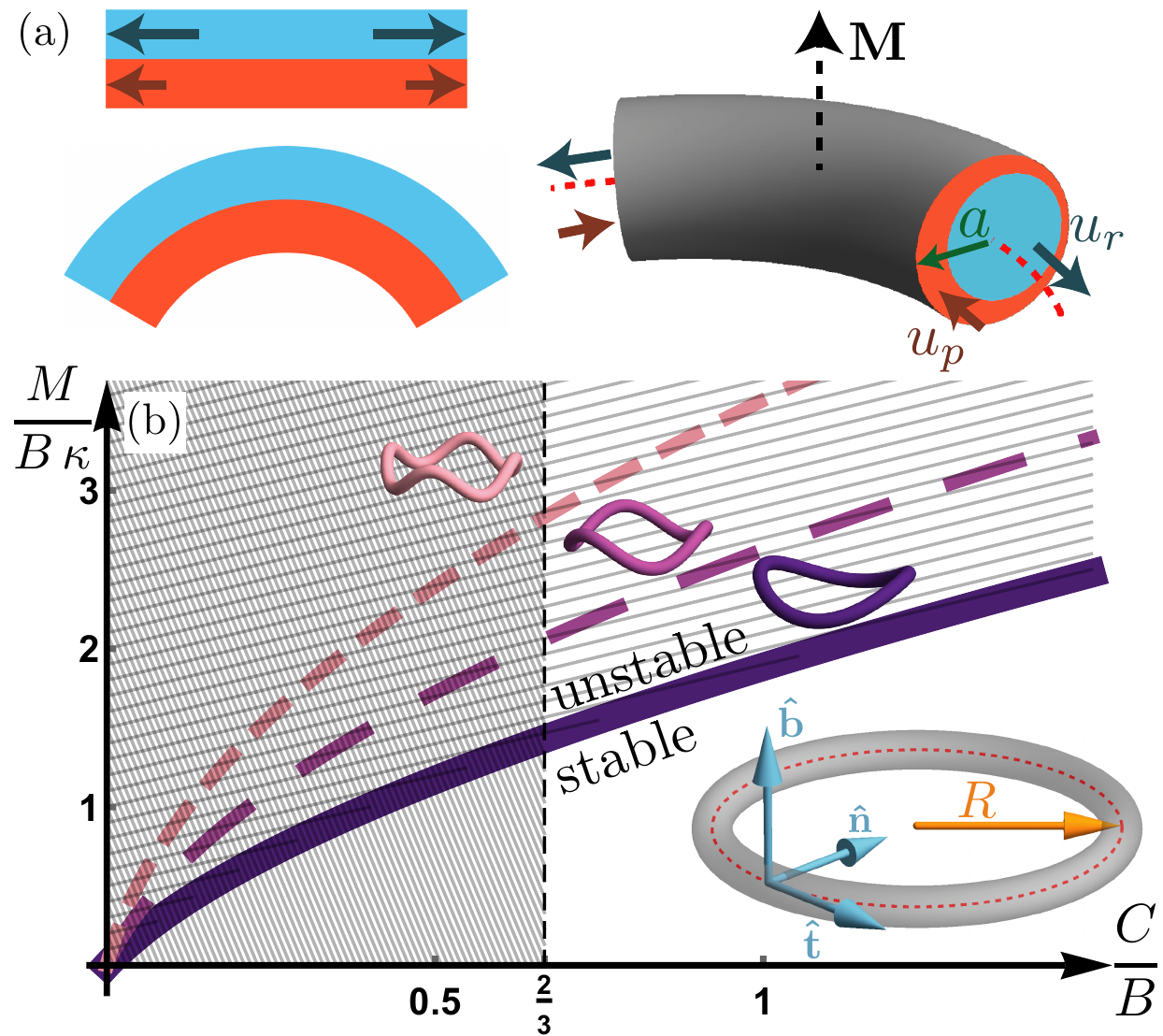}
\caption{\label{fig4} (a) Schematic of a bimetallic strip before (top left) and after (bottom left) heating. A slice through the cross-section of a phase-separated toroid is shown on the right with centerline (dashed red), cross-sectional radius $a$, polarized arrangement of solvent-rich (blue) and solvent-poor (orange) regions with corresponding strains $u_r$ and $u_p$, and the swelling moment $\mathbf{M}$. (b) Prediction of instability from linear stability analysis in terms of dimensionless measures of the swelling moment, $M/(B\, \kappa$), and the ring rigidity, $C/B$. The inset schematically shows the Frenet-Serret frame in an unperturbed ring, as well as the ``Pringling" and the next-two-lowest modes. Note that for uniform incompressible tori with a circular cross-section, elasticity theory dictates that $C/B \approx 2/3$.}
\end{figure}

As the boundary of the torus already consists of a collapsed-polymer layer, we expect that the solvent-poor region grows from this layer inward into the bulk, in a manner akin to heterogeneous nucleation. 
Furthermore, since a gel is a contiguous medium, the interface between solvent-rich and solvent-poor regions is laminated.
This interface frustrates the homogeneity of the polymer matrix and leads to a residual stress.
We therefore expect that the phase-coexistent state adopted by the gel will tend to minimize this inhomogeneity.
In the case of a sphere, the result is a solvent-poor skin of uniform thickness over the surface.
The non-constant curvature of the toroidal surface, however, leads to a skin of \emph{non-uniform} thickness.
Since the toroidal gel has higher ring curvature on its interior surface than on its exterior, we expect that a thicker polymer layer will form near the axis of revolution of the torus, as illustrated in the rightmost schematic in Fig.~\ref{fig4}(a).
This is indeed seen in experiment and is particularly clear at long-times, where the solvent-poor skin has clearly thickened and appears opaque, as shown in Figs.~\ref{fig3}(c,d).

To confirm our interpretations, we consider that, within the torus, a fraction $f$ of the gel is solvent poor and undergoes a volume change relative to its initial volume, $u_p < 0$. 
The remaining fraction ($1 - f$) of the gel is solvent-rich and increases its volume by a factor $u_r > 0$. 
The total volume constraint yields a ``lever rule'' $f u_p + (1 - f) u_r = 0$, which is a general feature in phase-separation with a conserved order parameter \cite{ChaikinLubensky,Callen1985}.
Using the Flory-Rehner theory of polymer gels \cite{sup-mat} and considering a cylindrical geometry, which amounts to neglecting ring curvature for now, we can determine equilibrium values for the strain $u_r$ and $u_p$ and the fraction $f$; from this, we confirm that phase-coexistent equilibria exist for temperatures $T$ above the LCST at constant volume and that the polymer volume fraction for the solvent-poor region is much larger than that of the solvent-rich region \cite{sup-mat}.
We then incorporate perturbatively the toroid's ring curvature on the phase-coexistence and find that it is favorable for the solvent-poor region to be thicker near the axis of revolution of the torus and thinner away from the axis \cite{sup-mat}, further confirming our previous assertions.
Interestingly, the resultant configuration is reminiscent of a bimetallic strip composed of two metals of differing thermal expansion coefficients that are laminated together, as illustrated in the leftmost schematic in Fig.~\ref{fig4}(a); under heating, the strip bends, increasing curvature due to the torque that results  from the differing thermal stresses in the two metals~\cite{Timoshenko1925}. 
In our case, the laminated coexistent phases of the gel have a similar stress differential.
We then focus on the ring-shape of the gel, ignore fine details of the cross-section, and develop a long-wavelength elastic model where the polymer matrix in the two laminated coexistent gel regions are each at \emph{fixed polymer volume fractions}.
Within this coarse-grained view of the gel, the net compressive stress, $\sigma$, exerted by the outer, solvent-poor shell on the inner, solvent-rich region is: $\sigma = E(u_{r} - u_{p})$, where $E$ is the gel's effective Young's modulus.

To describe buckling, we balance the stress $\sigma$ against the rigidity of the torus.
In general, toroidal bending is described by three-dimensional elasticity.
However, in our simplified model we treat the torus as an elastic rod defined by a circular centerline of length $L$.
This centerline is characterized at each point by its curvature $\kappa$ and torsion $\tau$, which are determined by the rotation rate of the Frenet-Serret frame; see inset in Fig.~\ref{fig4}(b). 
We consider an effective inextensible rod elastic free-energy $H$~\cite{LandauLifshitz1986}, in which the centerline degrees of freedom are encoded in changes in curvature $\Delta \kappa$ and changes in torsion $\Delta \tau$ at fixed length:
\begin{equation}
H = \int_0^L {\rm d}s\left(\frac{1}{2}\, B\, \Delta\kappa^2 + \frac{1}{2}\, C\, \Delta \tau^2 + \Delta \kappa \, \mathbf{\hat{b}} \cdot \mathbf{M} \right).
\label{eq:h}
\end{equation}
The first two terms in Eq.~(\ref{eq:h}) represent a rod with Hookean response to bending (and bending rigidity $B$) and twisting (and torsional rigidity $C$).
The third term in Eq.~(\ref{eq:h}) is associated to the swelling torque $\mathbf{M}$ acting on the centerline.
This model becomes strictly applicable in the limit $\xi \gg 1$. 
However, since the extensile rigidity remains much larger than the bending or torsional rigidities for significantly smaller $\xi$ \cite{dimitriyev}, it still applies down to the experimental values of $\xi$ where buckling is observed.

Right away, we see that our simple model indicates that the torus experiences swelling stresses that act to increase the ring curvature, reminiscent of the thermal stresses that bend bimetallic strips.
Owing to the relatively high energy cost of length changes, the torus is unable to attain a uniformly increased curvature whilst remaining planar, because any deformation that preserves both the length and winding number of a closed planar loop also leaves the integrated curvature for that loop unchanged \cite{Pressley2010}. 
To overcome this, the torus buckles out of the plane, which is what we observe experimentally.

To find the buckling threshold and modes, we perform a linear stability analysis \cite{Audoly2010,sup-mat} of Eq.~(\ref{eq:h}).
This analysis depends on two dimensionless numbers: the rigidity ratio $C/B$ and the stress ratio $M/(B \kappa$), where $M \equiv |\mathbf{M}|$.  
We find that the torus is unstable to buckling above a threshold value of $M/(B\kappa)$ at fixed $C/B$, as shown in Fig.~\ref{fig4}(b). 
This is seen in experiments, where tori with $\xi \lesssim 3$ do not buckle; see Fig.~\ref{fig3}(k,l) for a representative example.
Note that ``Pringling''  is the first of the buckling modes that is accessible upon increasing $M/(B \kappa$) at fixed $C/B$. 
For even larger $M/(B \kappa$), higher modes become unstable [Fig.~\ref{fig4}(b)].

Let us now estimate the quantities in Eq.~(\ref{eq:h}) and further compare to experiments.
The swelling torque $\mathbf{M} = - f x \pi a^2 \sigma \, \mathbf{\hat{b}}$ can be estimated as the cross-product of the lever arm $f x \, \mathbf{\hat{n}}$ with force $\pi a^2 \sigma \, \mathbf{\hat{t}}$. 
Here, $f$ is a good approximation of the fraction of the cross-sectional area occupied by the solvent-poor region, and $x$ is the center-of-area of the surface skin within the cross-section, which measures the imbalance of skin thickness due to surface curvature. 
Note that $x > 0$ because the shell is thicker closer to the axis of revolution of the torus.
To estimate the rigidities we consider that a uniform rod of circular cross-section radius $a$ has $B \approx \frac{1}{4} \pi a^4 E$ and $C/B \approx (1 + \nu)^{-1}$, with $\nu$ the Poisson ratio.
Crucially, since the gel is in the plateau period where the volume remains constant, we may regard it as \emph{rubber-like} and hence incompressible; thus we take $\nu \approx 1/2$ and $C/B \approx 2/3$.
We then find that $M/(B \kappa) \approx 4 f \xi$, where we have used that $x \approx a$, due to the highly polymer-dense region at the toroidal surface, and that $|u_p| \approx 1$, since this region contains very little solvent \cite{sup-mat}.
Theoretically, the buckling threshold for $C/B = 2/3$ is $M/(B\kappa) \approx 1.4$.
Considering that buckling is seen above $\xi \approx 3$, this implies that $f \approx 0.1$.
We can test this expectation by considering the ratio of deswollen to swollen gel volumes in the \emph{quasistatic} experiments (see Fig.~\ref{fig2}); in all cases, we obtain $f \approx 0.1$, consistent with the theoretical expectations.
Moreover, as $M/(B \kappa) \sim \xi$, the theoretical predictions of the buckling modes shown in Fig.~\ref{fig4}(b) relate well to the experiments.
Specifically, increasing $\xi$ in the experiments results in a transition from tori that are stable against buckling to ones that ``Pringle,'' and subsequently to tori that deform via more complicated shapes, which are reminiscent of the higher buckling modes predicted by the linear-stability analysis.

To further confirm that a swollen interior surrounded by a dense shell that is thicker near the axis of revolution results in buckling, we perform DPD simulations of toroidal shells having a nearly constant volume.
Since we model the toroidal shell by a 4-coordinated mesh of harmonic bonds \cite{sup-mat}, the curvature of the shell ensures that the effective rigidity of the portion closer to the axis of revolution is greater than the portion away from the axis, simulating the variable thickness observed in experiments. 
Remarkably, the simulations reproduce the ``Pringle\textsuperscript{TM}''-like shape seen experimentally, as shown for a torus with $\xi = 3.0$ in Figs.~\ref{fig3}(e,f).
Furthermore, the data can be fit to the hyperbolic paraboloid shape characteristic of Pringles \cite{sup-mat}.
Our simulations confirm that buckling is indeed related to the heterogeneous structure of our gels, in which a solvent-poor layer is forced to coexist with a solvent-rich bulk, and that in the process the volume of the gel remains essentially constant.
We also note that we also find modes other than ``Pringling.'' 
These are seen for higher values of $\xi$; an example is shown in Figs.~\ref{fig3}(i,j), which compares well with the experimental result shown in Figs.~\ref{fig3}(g,h). 
In addition, for sufficiently small $\xi$, no buckling is observed, consistent also with our experiments and theory, and buckling occurs only for $\xi \gtrsim 3$, also consistent with our experimental findings.

We have shown that rapidly heated tori composed of polymer gel can undergo constrained phase separation to form solvent-rich and solvent-poor regions and that the polarized arrangement of these regions within the torus can result in out-of-plane deformations. 
Our theoretical analysis also predicts that thin, curved pNIPAM gel rods would buckle when $f \gtrsim 0.35\, \xi^{-1}$, where in general $f$ is the volume fraction of solvent-poor gel and $\xi^{-1}=\kappa a$ is the product of rod curvature $\kappa$ and the tube radius.
While the shapes attained in our experiments are typical for rings that buckle due to mechanical instability, we emphasize that our results are entirely due to a \emph{thermodynamic instability}.
Thus, our work is suggestive of an ``extreme thermodynamics'' where shape actuation is achieved by passage through a phase transition.

This work was supported by the National Science Foundation (DMR-1609841, DMR-1207026, DMR-1255288). The work of PMG was also performed in part at the Aspen Center for Physics (NSF PHY-1607611). SVN is thankful to the NSF Graduate Research Fellowship program (DGE-1650044).

\bibliography{sources}

\end{document}